\documentclass{aastex}
\usepackage{emulateapj5}
\usepackage{epsfig}

\submitted{Accepted for publication in The Astrophysical Journal Letters, 2002 June 17}
\tighten

\shorttitle{High-Energy Emission from GRB 990104}
\shortauthors{Wren, Bertsch, \& Ritz}

\begin{document}

\title{Evidence for Postquiescent, High-Energy Emission\\ from Gamma-Ray Burst 990104}

\author{D. N. Wren}
\affil{Department of Physics, University of Maryland, College Park, MD 20742;\\Laboratory for High Energy Astrophysics, NASA/Goddard Space Flight Center,\\ Greenbelt, MD  20771}
\email{dnwren@milkyway.gsfc.nasa.gov}

\author{D. L. Bertsch, and S. Ritz}
\affil{Laboratory for High Energy Astrophysics, NASA/Goddard Space
Flight Center,\\ Greenbelt, MD  20771}
\email{dlb@egret.gsfc.nasa.gov}
\email{ritz@milkyway.gsfc.nasa.gov}

\begin{abstract}
It is well known that high-energy emission (MeV-GeV) has been observed in a number of gamma-ray bursts, and temporally-extended emission from lower energy gamma rays through radio wavelengths is well established.  An important observed characteristic of some bursts at low energy is quiescence: an initial emission followed by a quiet period before a second (postquiescent) emission.  Evidence for significant high-energy, postquiescent emission has been lacking.  Here we present evidence for high-energy emission, coincident with lower energy emission, from the postquiescent emission episode of the very bright and long burst, GRB 990104.  We show light curves and spectra that confirm emission above 50 MeV, approximately 152 seconds after the BATSE trigger and initial emission episode.  Between the initial emission episode and the main peak, seen at both low and high energy, there was a quiescent period of $\sim$100 s during which the burst was relatively quiet.  This burst was found as part of an ongoing search for high-energy emission in gamma-ray bursts using the EGRET fixed interval (32 s) accumulation spectra, which provide sensitivity to later, high-energy emission that is otherwise missed by the standard EGRET BATSE-triggered burst spectra.
\end{abstract}


\section{Introduction \label{1}}

Over a period of nine years, the Burst and Transient Source Experiment (BATSE) on the Compton Gamma Ray Observatory (CGRO) detected $\sim$2700 gamma-ray bursts\footnote[1]{Paciesas, W. S. et al. 2001, The Current BATSE Burst Catalog, available at http://cossc. gsfc.nasa.gov/batse/BATSE\_Ctlg/index.html} at a rate of approximately one per day \citep{meg92}.  Only a few are known for MeV-GeV emission \citep{snd92,snd95,hur94,kwo94,sfr94,sha94,som94,cat98,bri99,gon01}.  From two of these bursts, EGRET detected high-energy photons after low-energy emission ended: from GRB 940217 EGRET observed a $\sim$20 GeV photon, which arrived approximately 75 minutes after the cessation of low-energy emission \citep{hur94}; and from GRB 930131, EGRET observed MeV emission that continued for some time after the end of low-energy emission \citep{som94}.  Indeed, MeV emission from some bursts is coincident with low-energy emission, while from some bursts it is not \citep{din98}.  It is also known that bursts with a quiescent period between an initial emission episode (prequiescent) and a latter emission episode (postquiescent) are not uncommon \citep{ram01}.  However, significant high-energy spectral information has previously only been observed during prequiescent episodes (GRB 910503; Schneid et al. 1992), and high-energy spectral information from postquiescent emission has been lacking.\footnote[2]{Lower energy spectra have been observed by COMPTEL \citep{win92} }  This was due, in part, to data analysis choices that limited the emission search time intervals to those that were shorter than the burst quiescent time intervals.

Here we present light curves and high-energy spectra for the strong, post-quiescent emission episode of GRB 990104, which is noteworthy for several reasons: 
\begin{itemize}
\item with a 50--300 keV peak flux of $86.53 \pm 0.93$ photons cm$^{-2}$ s$^{-1}$  averaged over 64ms, it ranks fifth in brightness among BATSE catalog bursts;  
\item ·	its 50--300 keV fluence of ($1.67 \pm 0.11$) $\times$ 10$^{-4}$ ergs cm$^{-2}$ is greater than that of all but one burst in the BATSE catalog;
\item ·	its duration is greater than that of 96\% of catalog bursts for which duration is reported; and,  
\item ·	most significantly, as we show here, the light curves, count rates, and spectra indicate that there is a quiescent period of $\sim$100 seconds between the first emission episode and the most intense episode, during which the measured emission extends past 50 MeV.  This emission is coincident with low-energy emission described by the BATSE light curve, and provides the first measurable high-energy EGRET spectra from a postquiescent episode.  

\end{itemize}

\section{Time Profiles \label{2}}

\subsection{Burst Position}
During the burst, CGRO was oriented such that the EGRET spark chamber field of view was occulted by the Earth.  (For a description of the EGRET instrument see Thompson et al. 1993.)  In addition, the spark chamber was powered off, which was standard practice when the telescope was occulted.  However, two other components of the EGRET instrument, the Total Absorption Shower Counter (TASC) NaI calorimeter, and the anticoincidence scintillator dome, were omnidirectional, making burst detection possible.  The TASC and anticoincidence dome yield no positional information, so the direction for the burst was taken from the BATSE catalog.  This placed the burst at {\it l}$^{II}$ = $224.93\,^{\circ}\mathrm{}$, {\it b}$^{II}$ = $24.51\,^{\circ}$, which is $155\,^{\circ}$ from normal in spacecraft coordinates; therefore, the burst came from ``underneath'' the telescope.  We note that another CGRO instrument, COMPTEL, has no published listing of this burst.

\subsection{TASC Operation}
The TASC operated in two modes, with the normal mode accumulating spectra continuously over 32.768 s intervals.  When a burst trigger was received from BATSE, the TASC transitioned to burst mode, during which time spectra were taken for 1, 2, 4 and 16 s consecutive intervals.  These burst mode intervals, while useful for bursts in which emission came soon after a trigger, did not aid with spectrum measurements for longer bursts.  However, normal mode accumulation would resume when a 23 s burst mode accumulation was over, providing continuous accumulation of spectra in 32.768 s time bins.  Examination of the normal mode spectra revealed later high-energy emission for GRB 990104.

\subsection{Light Curves}
Continuous data of another sort was provided by EGRET's anticoincidence dome.  The primary task of the anticoincidence dome was to provide a means of vetoing signals from cosmic rays interacting with the telescope.  It also was sensitive to large fluxes of low energy gamma rays above $\sim$100 keV, allowing for the measurement of light curves with 0.256 s resolution.  In Figure 1, this anticoincidence count rate is shown between a BATSE 25--60 keV light curve (above), and TASC rate counts (below).  BATSE, with a resolution of 0.064 s, clearly has the most detailed curve, while the TASC rates have coarse 2.048 s time bins.  However, the TASC discriminators, with energy thresholds of about 1, 2.5, 7, and 20 MeV, provide information on the time profile in a higher energy range.  Of these, emission is visible in the $>$1, $>$2.5, and $>$7 MeV rates.  All three detector systems show a spiky rise at or near the BATSE trigger time, T, of 16:02:33.72 UT (57753.72 s), followed by a 100 s quiescent period before the main emission episode, which begins $\sim$152 seconds after the trigger (T+152 s).

\subsection{Burst Background}
Summing spectral counts provides another means of examining the activity of the burst.  Summing the raw TASC counts in a 1--200 MeV spectrum and dividing by the livetime gives an average count rate for that time interval.  Figure 2 is a plot of raw count rates for normal mode (32.768 s) spectra, covering a 720 s span.  There is a clear rise in rates at the time of the main emission, on top of a slow, steady decay in the background rate.  To define a background sample, we excluded the burst intervals in the span of (T - 5)--(T + 192) s, allowed for a one accumulation interval gap before and after the burst, and selected the next seven intervals both before and after the burst.  We fit a line to the background rates in these intervals [(T - 267)--(T - 37) s and (T + 224)--(T + 454) s], and derived estimated background rates in between.  Because the overall decay in background rates was smooth, the burst-time--derived background rates were insensitive to the choice of pre- and postburst background intervals.  Figure 3 shows the background subtracted normal mode count rates.  Comparison of Figures 1 and 3 shows a strong correlation between the time profiles of light curves and spectral counts.

We note that there may be a hint in the anticoincidence rates and lowest threshold TASC counters that there also is emission at T+84 s, but this not visible in the spectral rates in Figure 3, or in the BATSE light curve, and the weak signal disallowed a spectrum measurement.

\section{High-Energy Spectra \label{3}}
Spectrum measurements require TASC detector spectra to be converted to photon spectra through the use of a direction and energy-dependent response matrix.  These factors are computed with the CGRO mass model and EGS4 Monte Carlo code \citep{nel85}, which accounts for the detector geometry and intervening spacecraft material.

The background-subtracted rates in Figure 3 are a guide as to whether a given interval can yield a spectrum measurement.  Figures 4a and 4b show strong spectra during the two normal mode intervals corresponding to (T + 126)--(T + 159) s and (T + 159)--(T + 192) s, where spectral count rates are highest.  For the first of these, a single power law of the form {\it F(E) = $\alpha$(E/MeV)$^{-\beta}$} is fit over 1--20 MeV, with a resulting spectral index of $2.66 \pm 0.17$ and normalization constant of $0.64 \pm 0.12$ photons cm$^{-2}$ s$^{-1}$ MeV$^{-1}$.  Light curves indicate that the second interval covers more of the emission, which allows a single power-law fit over 1--100 MeV.  This gives an index of $2.52 \pm 0.03$ and normalization constant of $2.68 \pm 0.12$ photons cm$^{-2}$ s$^{-1}$ MeV$^{-1}$.  For comparison, when this second spectrum is fit over the same 1--20 MeV range as the first, the result is a constant of $2.64 \pm 0.12$ photons cm$^{-2}$ s$^{-1}$ MeV$^{-1}$ an index of $2.50 \pm 0.04$, consistent with the 1--100 MeV fit.  Though these normal mode spectra were strong, none of the four burst mode intervals had enough signal for spectrum measurements.  

The spectral characteristics of this burst are toward the high end of the $\beta$=1.7 to 3.7 distribution reported by \citet{cat98} for the 16 bursts they considered, while slightly harder than the TASC-only, $\beta$=$2.71 \pm 0.08$, of the much-studied GRB 990123 \citep{bri99}.

\section{Summary \label{5}}
We have observed high-energy emission from the post-quiescent emission episode of the very bright gamma-ray burst, GRB 990104.  Light curves over a wide energy range (25 keV -- 7 MeV) are consistent in showing that the main emission begins $\sim$152 seconds after the BATSE trigger.  This is confirmed by spectral count rates and two strong spectra, which cover the time interval (T + 126)--(T + 192) s.  The burst spectra are somewhat softer than most of the others that have previously been detected in the EGRET TASC, with spectral indices $2.66 \pm 0.17$ and $2.52 \pm 0.03$.  The first spectrum is strong enough for a fit from 1--20 MeV, and the second up through 100 MeV, with no high-energy cutoff observed.

It has been known for some time that an exciting question for the upcoming GLAST era is the behavior of the high-energy emission observed minutes or more after the initial emission.  The characteristics of GRB 990104 support the necessity of extended observations.

\acknowledgments
We gratefully acknowledge useful conversations with Scott Barthelmy, Neil Gehrels, Robert Hartman, Jay Norris, and David Thompson.

\clearpage

\begin{figure*}[tbp!]

\vspace{-5mm}
\hspace{15mm}
%\begin{center}
\psfig{file=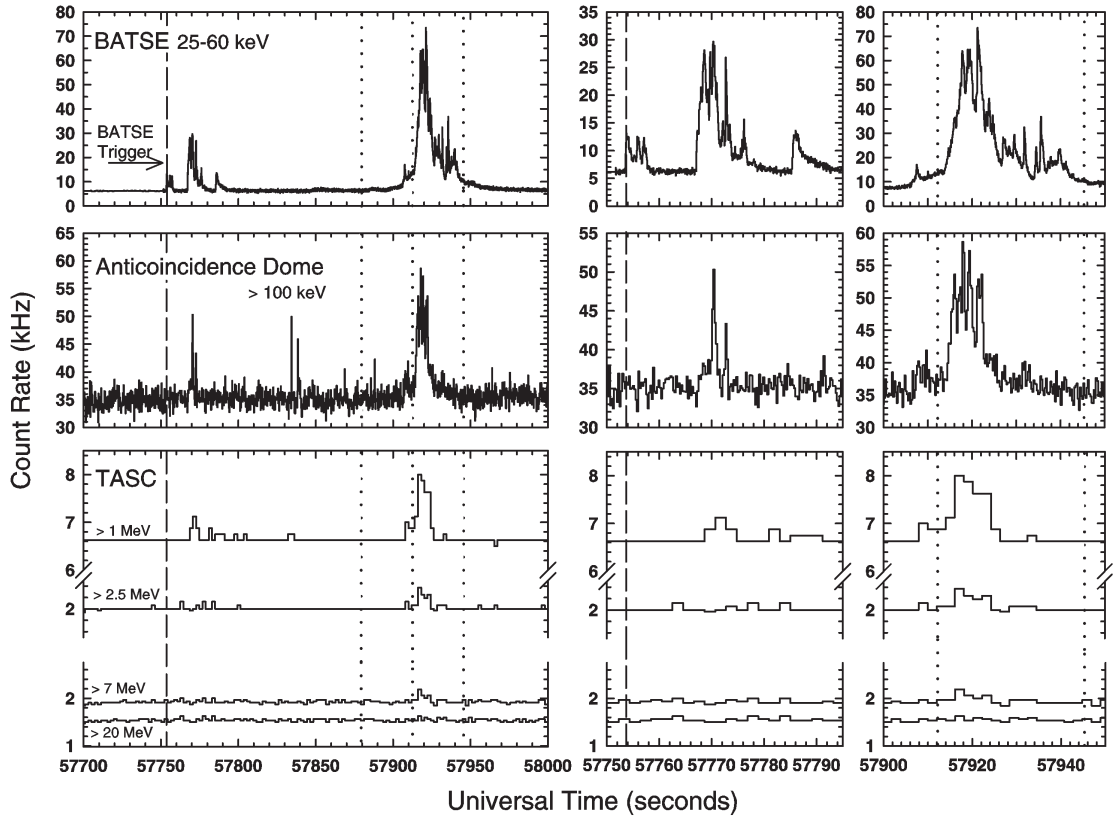,angle=270,width=6in}

\caption{Light curves from BATSE, EGRET's anticoincidence dome, and EGRET's TASC NaI calorimeter.  Plots on the right are expanded regions of the full light curves on the left.  The vertical dashed line indicates the time of the BATSE trigger.  The dotted lines mark the beginning and end of the two normal mode (32.768 s) spectra time bins shown in Figure 4.}
%\end{center}
\end{figure*}

\begin{figure*}[tbp!]

\begin{center}
\psfig{file=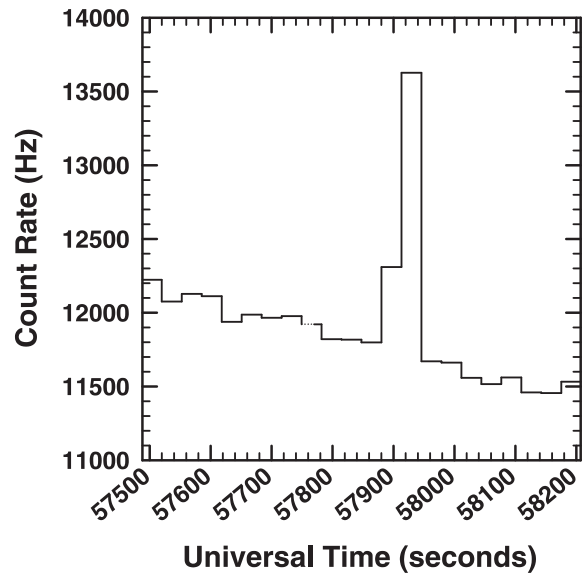,width=3in}
\caption{TASC raw count rates for normal mode (32.768 s) intervals.  A dotted line is drawn through the times when the TASC was in burst mode, and normal mode spectra were not being integrated.}
\end{center}
\end{figure*}

\begin{figure*}[tbp!]

\begin{center}
\psfig{file=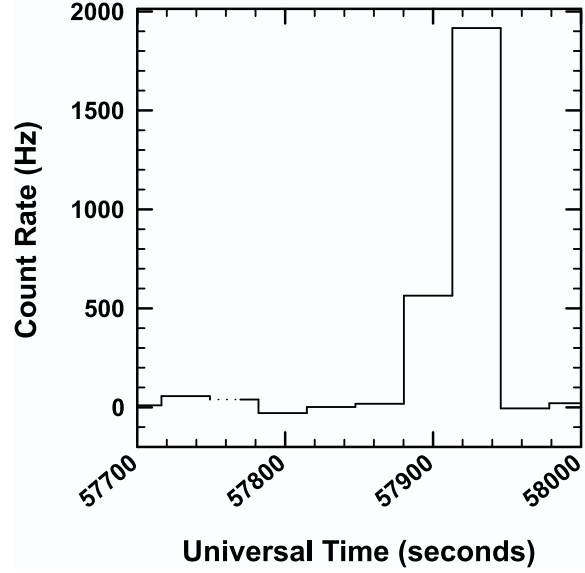,width=3in}
\caption{TASC background subtracted count rates for normal mode (32.768) intervals.  A dotted line is drawn through the time period for which the TASC was in burst mode, and normal mode spectra were not being integrated.}
\end{center}
\end{figure*}

\begin{figure*}[tbp!]
\begin{center}
\psfig{file=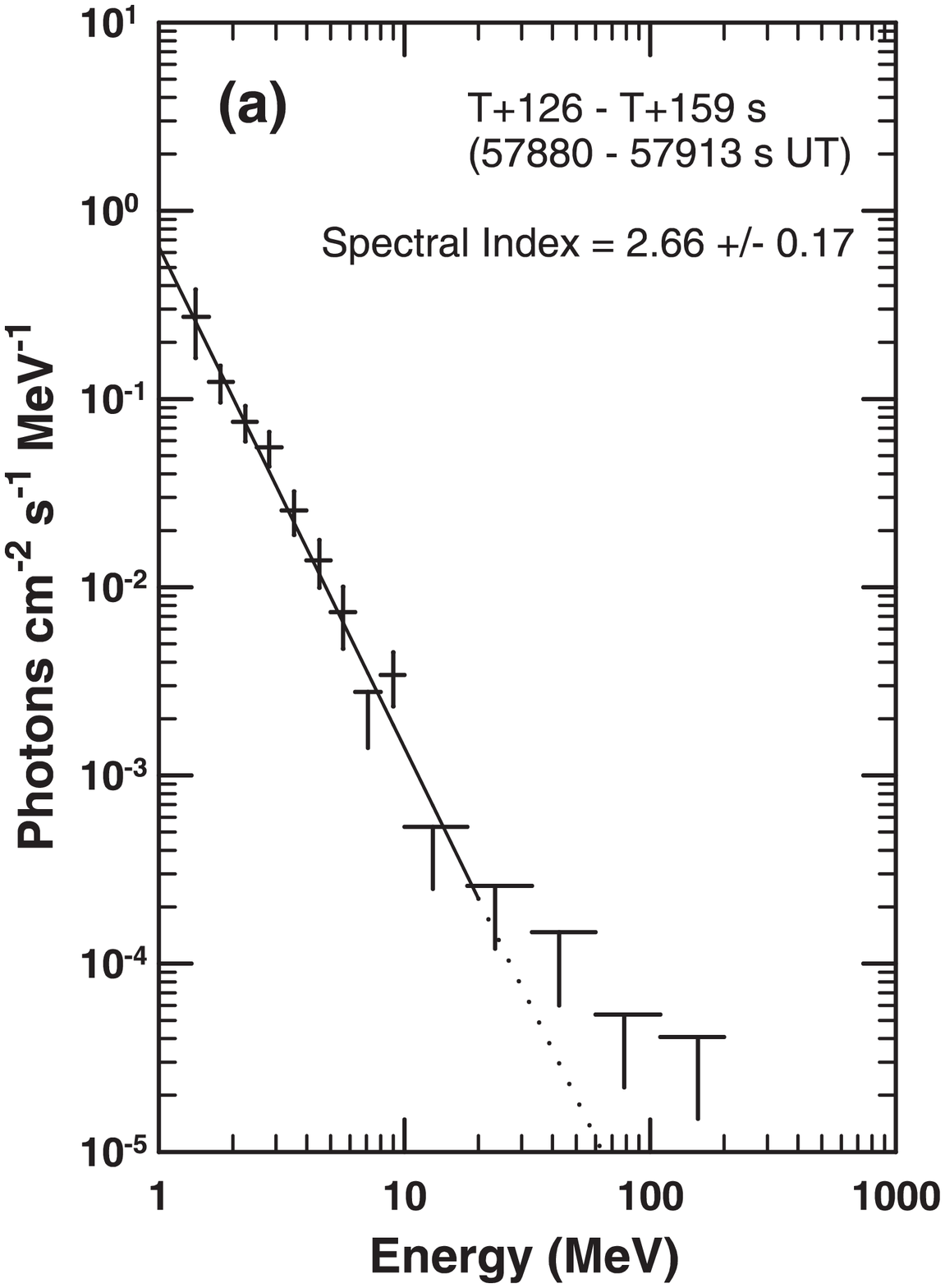,width=3.5in}
\psfig{file=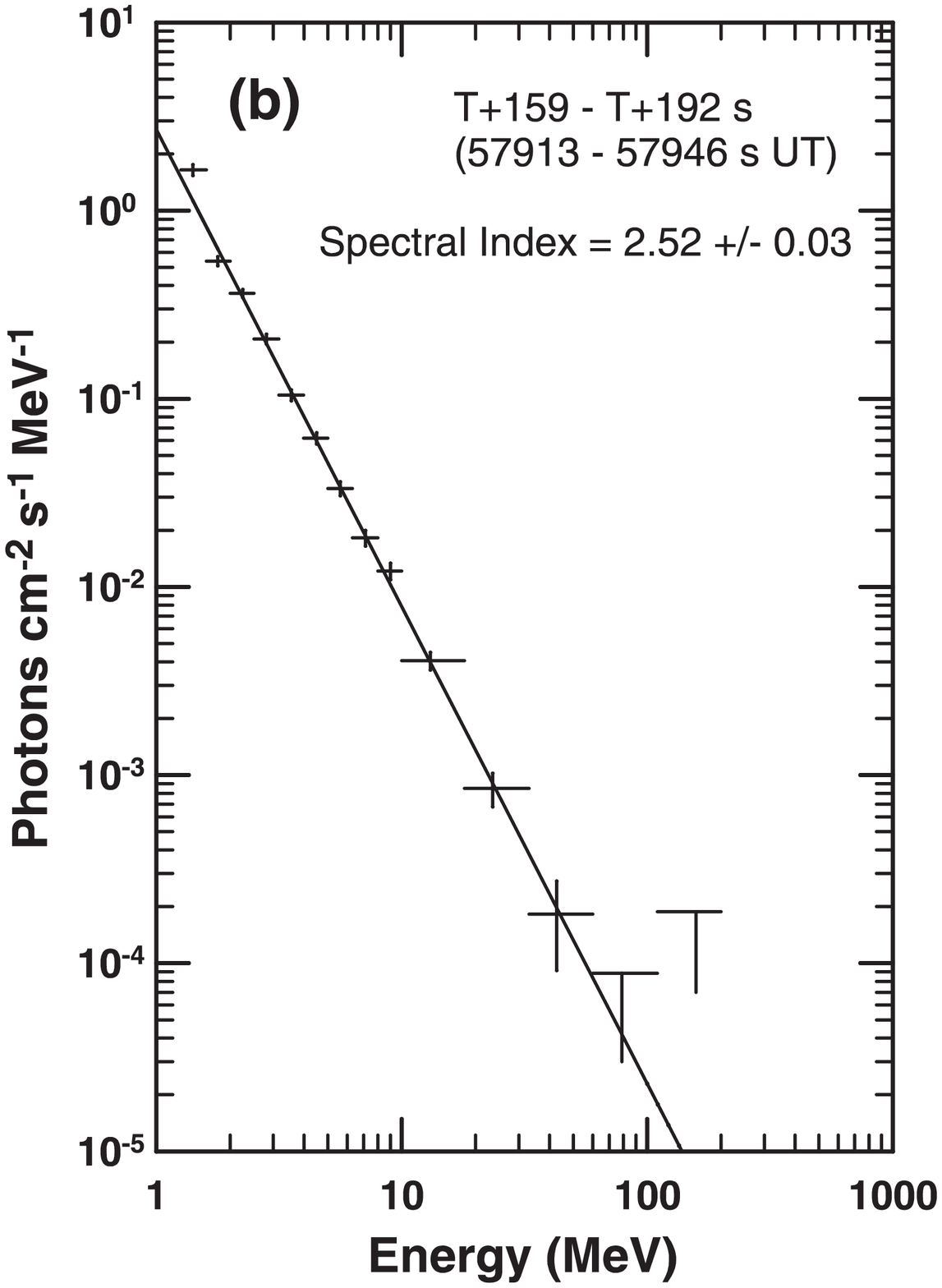,width=3.5in}
\caption{TASC photon spectra for two normal mode (32.768 s) intervals.  (a) Single power law fit over 1--20 MeV as a line passing through the data points.  This is for the normal mode (32.768) interval covering (T + 126)--(T + 159) s (57880--57913 s UT).  The dotted line is an extrapolation of that fit to higher energies.  The spectral index was measured to be $2.66 \pm 0.17$, with a normalization constant of $0.64 \pm 0.12$ photons cm$^{-2}$ s$^{-1}$. (b) Single power law fit over 1--100 MeV, covering the normal mode interval (T + 159)--(T + 192) s (57913--57946 s UT). The spectral index was measured to be $2.52 \pm 0.03$, with a normalization constant of $2.68 \pm 0.12$ photons cm$^{-2}$ s$^{-1}$.}
\end{center}
\end{figure*}

\clearpage

\end{document}